\documentclass[11pt,a4paper]{article}
\usepackage{jcappub}

\def\cha{{\it Chandra}}
\def\arcmin{^{'}}
\def\arcsec{^{''}}

\title{Willman 1: An X-ray shot in the dark with
  Chandra} 

\author[a]{D. Nieto} 
\author[a,b]{and N. Mirabal}

\affiliation[a]{Dpto. de F\'isica At\'omica, Molecular y Nuclear,
  Universidad Complutense de Madrid, Spain} 
\affiliation[b]{Ram\'on y Cajal Fellow}

\emailAdd{nieto@gae.ucm.es}
\emailAdd{mirabal@gae.ucm.es}

\abstract{The sterile neutrino is a weakly-interacting particle that
  emerges in the framework of certain extensions of standard particle
  physics and that fits naturally with the properties of a warm dark
  matter particle candidate. We present an analysis of a deep archival
  \cha\ observation of Willman 1, one of the darkest ultra-faint dwarf
  galaxies up to date, to exclude the presence of sterile neutrinos in
  the 1.6--16 keV mass range within 55 pc of its center down to the
  limiting flux of the observation. Spectral analysis of the
  \cha\ data fails to find any non-instrumental spectral feature
  possibly connected with the radiative decay of a dark matter
  particle. Accordingly, we establish upper bounds on the sterile
  neutrino parameter space and discuss it in the context of previous
  measurements.  Regarding the point source population, we identify a
  total of 26 sources within the central $5\arcmin$ to a limiting
  $0.5-2.0$ keV X-ray flux of $6\times 10^{-16}$ erg cm$^{-2}$
  s$^{-1}$.  While some of these sources could be formal members of
  Willman 1, we find no outstanding evidence for either an unusual
  population of bright X-ray sources or a densely populated cluster
  core. In fact, the entire X-ray population could be explained by
  background active galactic nuclei and/or foreground stars unrelated
  to Willman 1. Finally, possible associations of the X-ray point like
  population with optical sources from the {\it SDSS DR7} catalogue
  are presented.}


\arxivnumber{1003.3745}

\begin{document}

\maketitle

\flushbottom

\section{Introduction}

There is strong evidence for the existence of dark matter (DM) in the
Universe. In the concordance cosmological model (CCM), 83\% of the
mass density in the Universe cannot be explained with ordinary
baryonic matter and requires an additional non-baryonic
component \citep{Komatsu:2010a}. Without a doubt, understanding DM is one of the most
important topics of physics today \citep{Gaitskell:2004a,Bertone:2005a}.  In the
search for DM, three different approaches coexist: direct production
in collider experiments \citep{Kane:2008a}, direct detection through
scattering off ordinary matter \citep{Cerdeno:2010a}, and indirect detection
based on the search for secondary particles produced by the
annihilation or decay of DM particles \citep{Bertone:2005b}.

Within the last category, indirect searches for supersymmetric
particles at center-of-mass energies between 10 GeV and a few TeV from
pair annihilations or particle decays are being performed through
gamma-ray observations of astrophysical objects with high dark matter
density \citep{Aliu:2009a,Abdo:2010b}. While no significant dark matter signal has
been verified using gamma-ray photons as messenger particles
\citep{Anderson:2010a} some intriguing claims have emerged from 
measurements of the cosmic-rays positron fraction
\citep{Chang:2008a,Adriani:2009a} and electron spectrum 
in the same energy range \citep{Abdo:2009a,Aharonian:2009a}.

There are several theories offering DM particle candidates which can
annihilate into standard model particles and eventually produce
photons and leptons at energies above 10 GeV.  A neutralino in the
supersymmetric extensions of the standard model \citep{Haber:1985a} or the
Kaluza-Klein particle emerging from universal extra dimension theories
\citep{Bergstrom:2005a} are two well known examples of cold dark matter
(CDM) particle candidates. However, since the nature of the DM
particle remains uncertain, one must broaden the range of hypothesised
candidates/signatures taking in consideration options outside the
gamma-ray regime \citep{Feng:2010a}.  If we move to the X-ray band, a
possible candidate emerges in the form of the sterile
neutrino. Sterile neutrinos are weakly-interacting fermions which
arise as the right-handed counterparts of the standard neutrinos in
some extensions of the standard model of particles. The lightest of
these might lie in the keV range and would be compatible with a warm
dark matter (WDM) candidate \citep{Dodelson:1994a}. The sterile neutrino,
besides qualifying as a good WDM particle candidate and resolving the
neutrino mass problem, overcomes the puzzle of baryon asymmetry in
some of these extensions of the standard model \citep{Asaka:2005a} and may
mitigate some of the shortcomings of cold dark matter cosmologies
including the apparent dearth of dwarf galaxies around the Milky Way
\citep[but see][]{deNaray:2009a}.

Sterile neutrinos decay in a two-body final state composed by a
standard neutrino and a photon. Therefore, compact regions with
significant 
accumulations of sterile neutrinos could ``shine'' in X-rays, producing a
detectable X-ray flux line in the 0.1--100 keV energy range
\citep{Abazajian:2001a,Dolgov:2002a,Yuksel:2008a}. Ultimately, if 
the sterile neutrino is ever found
in collider experiments, a clear identification of this particle 
in dense astrophysical regions would have to follow 
\citep{Gelmini:2009a}. As we have discussed above,
an unidentified X-ray line could be an invaluable DM smoking gun.

Since we are dealing with a decay, the subsequent line flux is
directly proportional to the density of the DM region. Hence,
astrophysical objects with the highest DM density are the favoured
targets. Another condition that one should take into account when defining
the best targets is whether the expected X-ray signal would be
accompanied by any unrelated source of background that could eventually
curtain a sterile neutrino signal. Accordingly, 
the most prominent targets to date are the recently discovered ultrafaint
dwarf galaxies around the Milky Way \citep{Willman:2005a,Belokurov:2007a}. These
objects, whose kinematic properties are still under study
\citep{Simon:2010a}, could be the closest and densest dark matter structures
in the Local Group, making them excellent targets to conduct DM
searches \citep{Strigari:2008a,Simon:2010a}. Additionally, their low baryonic
content could minimise any external X-ray background. Nonetheless these
predictions must be verified through  careful study of the diffuse and
point-like X-ray emission coming from these objects.

Indirect searches for sterile neutrinos in dwarf galaxies have been
conducted in Fornax \citep{Boyarsky:2010a}, Ursa Minor \citep{Loewenstein:2009a},
Willman 1 \citep{Loewenstein:2010a}, and Segue 1 \citep{Mirabal:2010a}.  Here, we perform
such a search using a deep archival \cha\ observation of the enigmatic
Willman 1 \citep{Willman:2005a}, an ultra-faint object discovered in the
{\it Sloan Digital Sky Survey} ({\it SDSS}) and originally classified as a dwarf
galaxy \citep{Willman:2005a,Willman:2006a}. 
The arganization of the paper is as follows.  \S 2 describes the X-ray
observations. In \S 3 we discuss the diffuse component of Willman 1
and derive the parameter space for sterile neutrinos allowed by the
data. The X-ray point source population and spectral analyses are 
explained in \S
4. Summary and conclusions are presented in \S 5.

\section{Observations}

\begin{figure}
\hfil
\includegraphics[width=0.6\linewidth]{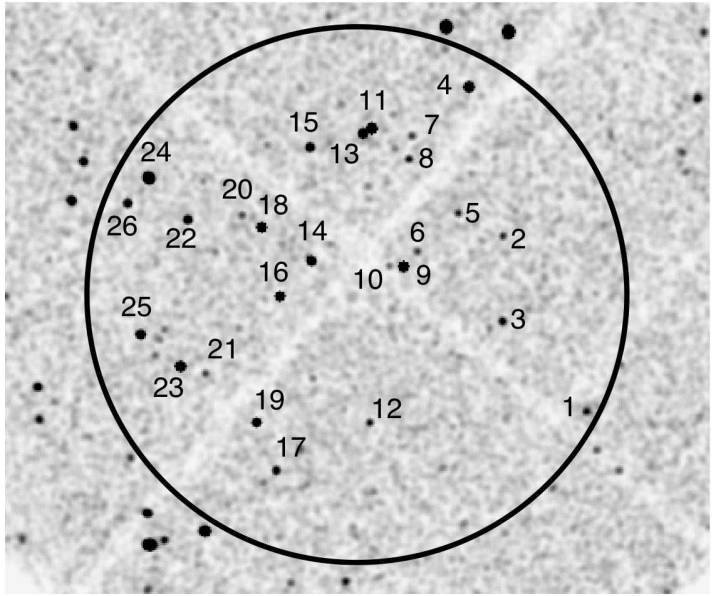}
\hfil
\caption{\cha\ ACIS-I image of Willman 1 in the 0.5--6.0 keV
energy range. 
The X-ray image has been smoothed with a
Gaussian kernel with radius $r_{k} = 2.5\arcsec$.
Identified sources
are marked following the labels adopted in Table~\ref{table_counts}.  
The circle is centred on RA. $10^{\rm h}49^{\rm m}22.\!^{\rm s}3$,
Dec. $+51^{\circ}03\arcmin03\arcsec$
and drawn with a 5$\arcmin$ radius. The half-light
radius corresponds to $r_{1/2} = 1.9\arcmin$.
Notice that 
the placement falls near the intersection of the chip gaps.
}
\label{xray_image}
\end{figure}

Willman 1 was observed by the {\it Chandra X-ray Observatory} for
102.75 ks on 2009 January 27--28 (\cha\ ObsID 10534) as described in
\citet{Loewenstein:2010a}. Observations were conducted with the Advanced CCD
Imaging Spectrometer (ACIS) in very faint ({\it VFAINT}) mode. Chips
0,1,2,3,6,7 were used with Willman 1 positioned near the ACIS-I
aimpoint on the ACIS-I3 chip.  Data reduction was performed using
standard
procedures\footnote{\url{http://cxc.harvard.edu/ciao/threads/aciscleanvf/}}
within the \cha\ Interactive Analysis of Observations (CIAO) software.
The level 1 event file was reprocessed to include grades 0,2,3,4, and
6. The total usable data is reduced to 100.68 ks after removing
periods of potential flares and elevated background.
Figure~\ref{xray_image} shows the resulting \cha\ X-ray (0.5--6.0 keV)
image. The X-ray image has been smoothed with a Gaussian kernel with
radius $r_{k} = 2.5\arcsec$.

\section{Diffuse emission and the search for a dark matter signature}

Based on the high mass-to-light ratio ($\sim 700$) inferred by
\citet{Martin:2007a}, Willman 1 has emerged as an attractive target for
indirect DM searches \citep{Strigari:2008a}. Yet, to date no DM signal has
been observed in sensitive observations of Willman 1 at energies $>
100$ MeV \citep{Aliu:2009a,Abdo:2010b}.  

In X-rays, one expect that the diffuse
emission in Willman 1 will be dominated either by an unresolved point
source population or by a complex gaseous component \citep{Hui:2009a}.
However, there is also a possibility that the DM halo of Willman 1
might produce a detectable X-ray flux via sterile neutrino decay. The
principal decay channel for sterile neutrinos is into three light
active neutrinos. Nonetheless, there is a potentially detectable
radiative decay channel in the X-ray band, whereby a lighter active
neutrino and a X-ray photon are produced with a slow mean decay on the
order of the lifetime of the Universe. The former is a two-body decay,
meaning that the resulting photon energy distribution is characterised
by a spectral line with a broadening due to the velocity dispersion of
the original sterile neutrino population.

In order to survey the diffuse emission, we excised all point sources
within the inner $5\arcmin$ of Willman 1.  Due to an unfortunate
placement of the object within the ACIS footprint, the diffuse
emission is missing a critical section of the inner core of Willman 1
and extends over four distinct ACIS-I CCDs with distinct gains (see
Figure~\ref{xray_image}). In order to avoid significant intrachip
variations, we excluded 32 pixels to each side of the CCD
edges. Visual examination of the resulting X-ray image does not reveal
prominent diffuse emission that would give away a rich globular
cluster core.  Given the intrachip gain fluctuations across the CCDs,
it is not straightforward to ascertain the significance of the counts
within the inner $5\arcmin$ of Willman 1.  We note however that the
limited signal-to-noise ratio and placement across four CCDs cannot
ensure that the detected counts correspond to diffuse emission
specific to Willman 1. Rather the counts might result from Poisson
fluctuations in the background count rate.

\begin{figure}
\hfil
\includegraphics[width=0.7\linewidth]{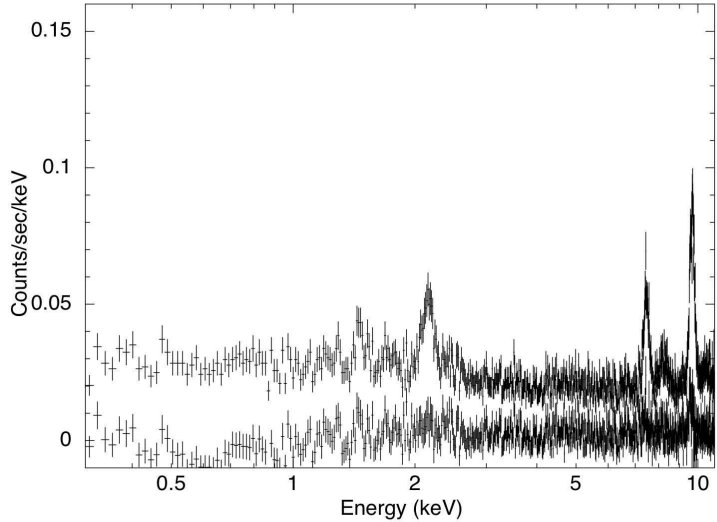}
\hfil
\caption{\cha\ ACIS-I spectrum extracted from 
a $5\arcmin$ radius circle centred on Willman 1 as described in the text.
{\it Top:} Data prior to
removal of the background. Instrumental line features 
include Si K $\alpha$ (1.74 keV),
Au M $\alpha,\beta$ (2.1--2.2 keV), Ni K $\alpha$ (7.47 keV), and Au L
$\alpha$ (9.67 keV).  
{\it Bottom:} Spectrum after background subtraction.
}
\label{acis_spectrum}
\end{figure}

On the other hand, it is possible to derive upper limits on the
diffuse emission after subtracting the background contribution at this
position. The spectrum of the diffuse emission was rendered from a
$5\arcmin$ radius circle at the nominal position of Willman 1 of
RA. $10^{\rm h}49^{\rm m}22.\!^{\rm s}3$,
Dec. $+51^{\circ}03\arcmin03\arcsec$ and corrected with the standard
reprojected blank-sky background.\footnote{\url{http://cxc.harvard.edu/ciao/threads/acisbackground/index.sl.html}}
The final step consisted in a renormalization of the background
spectrum to match the 9.0--12.0 keV count rate of the Willman 1
exposure.  Figure~\ref{acis_spectrum} shows the spectrum
before and after subtraction of the background.  Prior to subtraction,
the spectrum is dominated by prominent instrumental line features that
originate from fluorescence of material in the telescope including Si
K $\alpha$ (1.74 keV), Au M $\alpha,\beta$ (2.1--2.2 keV), Ni K
$\alpha$ (7.47 keV), and Au L $\alpha$ (9.67 keV). The instrumental
features are largely removed after subtracting the background.

In order to establish upper limits to the emission of DM in
the energy range covered by the observation, we fit the spectrum in
steps of 0.1 keV with the appropriate Gaussian width $\sigma$ required
to match the spectral resolution at each step.  This emulates the
procedure outlined in previous analyses of \cha\ data
\citep{Mirabal:2003a,RiemerSorensen:2006a}.  We next proceeded to derive limits coming
from a DM halo composed of sterile neutrino particles with
rest mass 1.6 keV $< m_{\nu} < 16.0$ keV as a function of mixing angle
$\theta$ adopting the formalism of \citet{Loewenstein:2010a},
\begin{equation}
F_{line}=5.15 \sin^2\theta \left({m_s \over keV}\right)^4f_sM_7d^{-2}_{100} cm^{-2} s^{-1}
\end{equation}
where sterile neutrinos with mass $m_s$ produce photons at a given
line energy $E_{line} = m_s/2$, $M_7$ is the projected DM
mass of Willman 1 in units of $10^{7} M_{\odot}$, $f_s$ is the
fraction of DM in sterile neutrino form and $d_{100}$ is the
distance to Willman 1 in units of 100 kpc. For the actual
calculations, we assume that the DM is composed by sterile
neutrinos only ($f_s = 1$), a heliocentric distance $d_{100} = 0.38$
\citep{Willman:2006a} and a projected DM mass $M_7 = 0.2$
\citep{Loewenstein:2010a}.  Figure~\ref{parameter_space} shows the resulting sterile
neutrino parameter space ruled out by the \cha\ observation. Except
for some minor differences, our results for the parameter space are
consistent with the results originally reported by \citet{Loewenstein:2010a}.

\begin{figure}
\centerline{
\hfil
\includegraphics[width=0.7\linewidth]{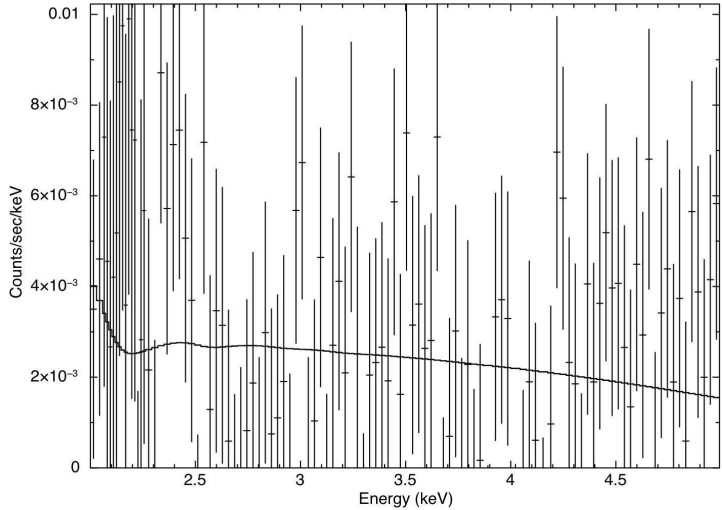}
\hfil
}
\caption{\cha\ ACIS spectrum of the diffuse component of
Willman 1 in the 2.0-5.0 keV energy range
and best-fitting absorbed power-law model as 
described in the text.
}
\label{acis_spectrum_zoom}
\end{figure}

However, we differ from \citet{Loewenstein:2010a} in that we find no evidence for a
purported line detection at 2.51 keV with a line flux $f_{line} = 3.53
\times 10^{-6}$ photons cm$^{-2}$ s$^{-1}$.  Figure~\ref{acis_spectrum_zoom}
shows the background-subtracted ACIS spectrum in the 2.0--5.0 keV
energy region.  An absorbed power-law model fixed at the Galactic H I
column density $N_{\rm H}$ = $1.2 \times 10^{20}$ cm$^{-2}$ and photon
index $\Gamma =1.0 \pm 0.6$ appears to accommodate the data
($\chi^{2}_{\nu} = 0.86$).  There is an excess around 2.3--2.5 keV,
but the inclusion of a Gaussian line with the properties reported in
\citet{Loewenstein:2010a} does not improve the fit ($\chi^{2}_{\nu} = 0.85$).
Leaving the power-law index to be a free parameter and allowing the
Gaussian line centroid wander in the 2.0--3.0 keV energy yields an
unrealistic $\Gamma = -1.2 \pm 0.7$ with a Gaussian line centred at
2.2 keV. The latter is most likely due to the instrumental Au M
$\alpha,\beta$ (2.1--2.2 keV) line and highlights the difficulties
with reported line detections in and around this region.  Thus, while
we cannot rule out weak emission between 2.3 and 2.5 keV, we find that
a power-law model provides a satisfactory fit without the need for the
Gaussian line reported by \citet{Loewenstein:2010a}.

Upon closer inspection of Figure 1 in \citet{Loewenstein:2010a} we notice spectral
residuals in the vicinity of the aforementioned instrumental lines of
Au M $\alpha,\beta$ (2.1--2.2 keV), Ni K $\alpha$ (7.47 keV), and Au L
$\alpha$ (9.67 keV). We face similar issues in our analysis that seem
entirely consistent with a diffuse emission signal heavily dominated
by background spectrum \citep[see also][]{Boyarsky:2010a}.  In such
circumstances, it is possible that a deficient subtraction of the Au M
$\alpha,\beta$ (2.1--2.2 keV) instrumental line or calibration issues
at $E < 2.3$ keV will start to carve the silhouette of a spectral line
around 2.4--2.5 keV that might result in the finding reported in
\citet{Loewenstein:2010a}.  In addition, it is difficult to discard that intrachip
gain variations may conspire to mimic an emission line. It is
important to recall that spurious X-ray line detections have a curious
history in the literature \citep{Sako:2005a,Vaughan:2008a}. Henceforth, one must
be careful when dealing with marginal detections \citep{Protassov:2002a}.

Even if one wants to advocate the reality of a line at 2.51 keV, a
direct DM connection suffers from a fatal flaw in that it
falls at a location matching the rest-frame of helium-like sulphur ion
S XV $\alpha$ located at 2.45 keV \citep{Raymond:1977a}.  This astrophysical
line is routinely detected in plasmas with temperatures $\sim
10^{6-7}$ K \citep{Hwang:2004a,Henley:2008a}.  Consequently, it seems more
plausible that one is detecting sulphur emission from a hot supernova
remnant or a highly-ionised wind region either at Willman 1 itself or
at an intervening location to the object, rather than the exotic
signature of DM.

\begin{figure}
\centerline{
\hfil
\includegraphics[width=0.5\linewidth]{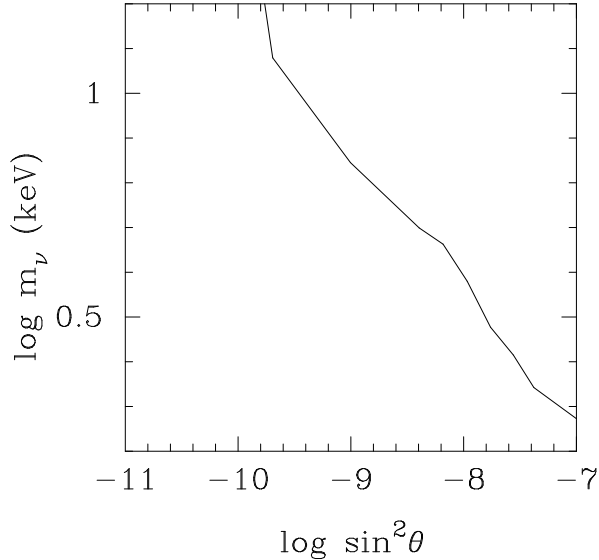}
\hfil
}
\caption{Parameter space constraints for 
the sterile neutrino. The region to the right of
the contour is ruled out by the \cha\ observation of Willman 1. 
}
\label{parameter_space}
\end{figure}

\section{X-ray point source population}

Apart from allowing access to the extended emission, the
\cha\ resolution is sufficiently detailed to study the X-ray point
source population in the field of Willman 1. For the purposes of this
work, we searched for point sources within the central $5\arcmin$ of
Willman 1 using the CIAO tool {\it celldetect}.  This corresponds to a
physical size of 55 pc assuming a distance of 38 kpc.  This region
encompasses the half-light radius of Willman 1 estimated to be
$r_{1/2}$ = $1.9\arcmin$ \citep{Willman:2006a}.  Figure~\ref{xray_image}
shows all 26 detections with more than 15 net counts in the 0.5--6.0
keV energy range, which translates into a luminosity limit for
point-like sources of $2.7\times 10^{32}$ erg s$^{-1}$. Ten of the
sources are on the I1 chip, seven on the I2 chip, six on the I0 chip,
and three on the I3 chip.

\begin{table*}
\begin{center}
\scriptsize
\begin{tabular}{ccccccc}
\hline
 & Name & R.A. & Dec. &\multicolumn{3}{c}{Net Counts} \\
Source Label & & (J2000.0) & (J2000.0) &
0.5 -- 1.5 keV &  0.5 -- 4.5 keV & 1.5 -- 6.0 keV \\
\hline
1  & CXOU J104855.4+510053 & 10 48 55.4 & +51 00 53.0 & 1$\pm 1$ &
14$\pm 4$   &  20$\pm 5$ \\
2 & CXOU J104905.1+510409 &10 49 05.1 & +51 04 09.6 & 1$\pm 1$ &
 16$\pm 4$    & 20$\pm 5$       \\
3 & CXOU J104905.2+510234 & 10 49 05.2 & +51 02 34.5 & 15$\pm 4$    &
29$\pm 5$    & 17$\pm 4$      \\
4 & CXOU J104909.1+510657 & 10 49 09.1 & +51 06 57.1 & 94$\pm 10$    &
 171$\pm 13$  & 83$\pm 9$       \\
5 & CXOU J104910.3+510435 & 10 49 10.3 & +51 04 35.8 & 15$\pm 4$    &
 20$\pm 5$    & 5$\pm 2$     \\
6 & CXOU J104915.1+510353 & 10 49 15.1 & +51 03 53.2 & 6$\pm 3$    &
14$\pm 4$    & 9$\pm 3$       \\
7 & CXOU J104915.9+510602 & 10 49 15.9 & +51 06 02.2 & 9$\pm 3$    &
 17$\pm 4$    & 8$\pm 3$      \\
8 & CXOU J104916.2+510536 & 10 49 16.2 & +51 05 36.6 & 18$\pm 4$    &
 26$\pm 5$    &  10$\pm 3$     \\
9  & CXOU J104916.8+510335 & 10 49 16.8 & +51 03 35.6 & 89$\pm 9$    &
 161$\pm 13$  & 83$\pm 9$       \\
10  & CXOU J104918.5+510337 & 10 49 18.5 & +51 03 37.0 & 7$\pm 3$    & 
12$\pm 4$ & 8$\pm 3$       \\
11 & CXOU J104920.6+510610 & 10 49 20.6 & +51 06 10.6 & 183$\pm 14$ &
 291$\pm 17$   & 124$\pm 11$      \\
12 & CXOU J104920.8+510041 & 10 49 20.8 & +51 00 41.3 & 15$\pm 4$    &
 22$\pm 5$    & 10$\pm 3$       \\
13 & CXOU J104921.6+510605 & 10 49 21.6 & +51 06 05.1 & 46$\pm 7$    &
81$\pm 9$     & 39$\pm 6$  \\
14 & CXOU J104927.7+510341 & 10 49 27.7 & +51 03 41.5 & 27$\pm 5$ &
 66$\pm 8$    & 39$\pm 6$    \\
15 & CXOU J104927.8+510549 & 10 49 27.8 & +51 05 49.3 & 28$\pm 5$ &
55$\pm 7$    & 33$\pm 6$  \\
16  & CXOU J104931.4+510302 & 10 49 31.4 & +51 03 02.5 & 58$\pm 8$    &
100$\pm 10$   & 47$\pm 7$      \\
17 & CXOU J104931.8+505947 & 10 49 31.8 & +50 59 47.0 & 12$\pm 4$    &
 33$\pm 6$    & 24$\pm 5$       \\
18 & CXOU J104933.6+510420 & 10 49 33.6 & +51 04 20.0 & 48$\pm 7$    &
 112$\pm 11$    & 70$\pm 8$      \\
19 & CXOU J104934.1+510041 & 10 49 34.1 & +51 00 41.1 & 28$\pm 5$    &
 61$\pm 8$    & 39$\pm 6$      \\
20 & CXOU J104935.9+510433 & 10 49 35.9 & +51 04 33.2 & 4$\pm 2$    &
 13$\pm 4$    & 11$\pm 3$      \\
21  & CXOU J104940.2+510136 & 10 49 40.2 & +51 01 36.0 & 4$\pm 2$    &
13$\pm 4$    & 14$\pm 4$       \\
22 & CXOU J104942.3+510428 & 10 49 42.3 & +51 04 28.1 & 36$\pm 6$   &
 57$\pm 8$   & 24$\pm 5$     \\
23 & CXOU J104943.1+510144 & 10 49 43.1 & +51 01 44.1 & 34$\pm 6$    &
 178$\pm 13$    &  157$\pm 13$     \\
24  & CXOU J104946.9+510514 & 10 49 46.9 & +51 05 14.7 & 154$\pm 12$
   & 316$\pm 18$   &   193$\pm 14$    \\
25 & CXOU J104947.8+510219 & 10 49 47.8 & +51 02 19.7 & 31$\pm 6$    &
78$\pm 9$   & 57$\pm 8$       \\
26  & CXOU J104949.4+510446 & 10 49 49.4 & +51 04 46.8 & 26$\pm 5$    &
 43$\pm 7$    &   19$\pm 4$      \\
\hline
\end{tabular}
\normalsize
\caption{X-ray sources identified within the central $5\arcmin$ of
  Willman 1. X-ray counts were extracted within a $2.5\arcsec$ radius
  circles centred on the source location.}
\label{table_counts}
\end{center}
\end{table*}

X-ray counts were extracted within a $2.5\arcsec$ radius circles
centred on the source location.  The counts were then separated in
three different energy ranges: a {\it soft} band (0.5--1.5 keV), a
{\it medium} band (0.5-4.5 keV), and a {\it hard} band (1.5-6.0
keV). Finally, we estimated net counts using source-free background
regions in the corresponding individual chip. The source labels,
positions, and net counts in three separate energy bands are listed in
Table~\ref{table_counts}.  A complementary search for optical
counterparts was performed cross-checking source positions with {\it
  SDSS DR7} catalogue \citep{Abazajian:2009a}. The association required the
angular distance between \cha\ and {\it SDSS} catalogue sources to be
less than $2.5\arcsec$. This criterion produces 10 optical matches out
of the total sample of 26 sources. According to {\it SDSS} object
classification, 3 are categorized as stars and 3 as galaxies. The
remainder 4 matches are listed as part of the {\it NBCK} catalogue of
photometrically selected quasar candidates \citep{Richards:2009a}.  Sources
with optical matches are summarised in
Table~\ref{table_associations}. It is important to emphasise that {\it SDSS}
optical sources have been tentatively labelled based on colours and
that no reliable spectroscopic classifications were available at the
time of this writing.

\begin{table*}
\scriptsize
\begin{center}
\begin{tabular}{cccccc}
\hline
 & Name & Optical match & Offset ($\arcsec$) & Classification \textdagger & Redshift \textdaggerdbl \\
Source Label \\
\hline
4 & CXOU J104909.1+510657 & SDSS J104909.10+510657.0 & 0.1 & Star & - \\
8 & CXOU J104916.2+510536 & SDSS J104916.19+510536.6 & 0.1 & Galaxy & - \\
11 & CXOU J104920.6+510610 & SDSS J104920.62+510610.5 & 0.2 & Star & - \\
12 & CXOU J104920.8+510041 & SDSS J104920.85+510041.3 & 0.5 & QSO \textasteriskcentered & 2.8 \\
13 & CXOU J104921.6+510605 & SDSS J104921.64+510605.1 & 0.4 & QSO \textasteriskcentered & 1.6 \\
16  & CXOU J104931.4+510302 & SDSS J104931.40+510302.6 & 0.2 & QSO \textasteriskcentered & 2.1 \\
17 & CXOU J104931.8+505947 & SDSS J104931.79+505946.9 & 0.1 & Galaxy & - \\
24  & CXOU J104946.9+510514 & SDSS J104946.90+510514.3 & 0.4 & QSO \textasteriskcentered & 0.8 \\
25 & CXOU J104947.8+510219 & SDSS J104947.85+510220.0 & 0.6 & Galaxy & - \\
26  & CXOU J104949.4+510446 & SDSS J104949.36+510446.6 & 0.3 & Star & - \\
\hline
\end{tabular}
\normalsize
\caption{X-ray sources with optical matches within the central
  $5\arcmin$ of Willman 1. \textdagger\ According to
  \citet{Abazajian:2009a}. \textdaggerdbl\ Photometric redshift from
  \citet{Richards:2009a}. \textasteriskcentered\ Photometric classification
  from \citet{Richards:2009a}}
\label{table_associations}
\end{center}
\end{table*}

The principal obstacle in sorting out whether any individual point source 
in the field represents a {\it bona fide} member of Willman 1 is the
inescapable presence of background active galactic nuclei (AGN) 
and galaxies in the field \citep{Giacconi:2000a,Luo:2008a}. It turns out to be extremely 
difficult to distinguish the source identity whenever 
the number of predicted background AGN matches the observed number of
sources in the field \citep{Ramsay:2006a}.   
In this instance,
a detection limit of 15 net counts in the 0.5--6 keV energy
range corresponds to a limiting flux $6\times 10^{-16}$ erg cm$^{-2}$ s$^{-1}$
 in the 0.5--2.0 keV band, adopting an X-ray power spectral index 
$\Gamma = 1.4$ and a Galactic H~I 
column density $N_{\rm H}$ =
$1.2 \times 10^{20}$ cm$^{-2}$. 
Using the latest AGN counts from the \cha\ Deep Field-South (CDFS) 
reported by
\citet{Luo:2008a}, we expect an average of $24 \pm 3$ background AGN for a
field of this size 
to our limiting flux in the 0.5--2.0 keV band. 
Our detection of 26 sources does not exceed significantly the 
CDFS prediction. 
As a result,   
the majority of sources found are most likely background AGN.

\begin{figure}
\hfil
\includegraphics[width=0.5\linewidth]{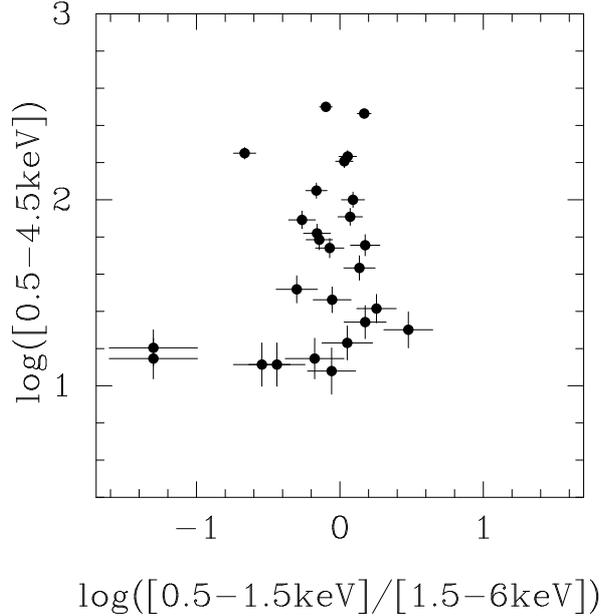}
\hfil
\caption{X-ray colour-magnitude diagram 
for 26 sources detected within the inner 5$\arcmin$  of Willman 
1.}
\label{color_magnitude}
\end{figure}

Since this result relies on the adopted background source count, 
we proceeded to evaluate the variability and
colour of the sample. The presence of binary systems might be exposed
through the detection of strong variability \citep{Heinke:2003a}.  According
to the CIAO tool {\it glvary}, there are no definite variables among
the point sources. Similarly, the location of a source in an X-ray
colour magnitude diagram (CMD) might be a powerful way to infer the
presence of binary systems in a field \citep{Grindlay:2001a}. In particular,
certain CVs/LMXBs will depart from AGN and become outliers in the
colour distribution. In Figure~\ref{color_magnitude} we plot the logarithm of
the 0.5--4.5 keV counts versus the logarithm of the ratio of 0.5--1.5
keV and 1.5--6.0 keV net counts. Note that the majority of the sources
tend to gather around the center of the diagram where it is difficult
to tell binary systems and AGN apart. However, the void in the upper
right corner of the diagram probably rules out the presence of LMXBs
in quiescence \citep{Heinke:2003a}. The two hard sources to the left of the
diagram could indicate heavily obscured AGN but the reduced number of
counts prevents a formal identification.  Therefore, the colour
analysis fails to discover any apparent binary system in the field as
expected by our estimation.

Additional clues about the nature of the point like population might
be gained from the examination of the actual spectra of the sources.
For this reason, spectra were generated for all the sources with 50 or
more detected counts in the 0.5--6.0 keV energy band.  This criterion
leaves 13 out of the total sample of 26 sources.  The counts for each
source were grouped using the CIAO tool {\it dmgroup} such that there
are 10 counts per bin. By default, we adopted a power-law model
spectral fit within XSPEC \citep{Arnaud:1996a} with the absorption fixed at
the Galactic value $N_{\rm H}$ = $1.2 \times 10^{20}$ cm$^{-2}$. Power
law fits with indices $1.2 < \Gamma < 2$ are acceptable
($\chi^{2}_{\nu} < 1.5$) to all sources, except source 14.  For the
latter a blackbody or models with substantial internal absorption
provide better fits.  However, it is difficult to make a final
spectral classification for that source. Overall, the spectral results
are consistent with the properties of background AGN in such a
field. We note that there is also a non-negligible probability that a
handful of coronal emitting stars in the foreground could be
superposed by chance along the line of sight \citep{Haggard:2009a}.

Last but not least, the number of expected X-ray binary systems in
Willman 1 can be estimated theoretically.  We computed the so-called
encounter rate following \citet{Verbunt:2003a} and considering a core radius
$r_{c}$ = 1.5$\arcmin$ and density $\rho_{0} \sim 1 \times 10^{-2}
L_{\odot}/pc^{-3}$ \citep{Martin:2007a}.  Applying the normalisation in
\citet{Pooley:2003a} we derive a very low probability estimate ($p \sim
\mathcal{O}(10^{-7}) $) of finding X-ray binary systems at a
luminosity over $4 \times 10^{30} erg~s^{-1}$. Given that our
luminosity limit falls two orders of magnitude above, the probability
estimate must be considered as an upper limit and therefore one would
not expect to detect any X-ray binary system belonging to Willman 1 in
this observation. Nevertheless the assumptions made in order to infer
the binary rate are based on the daring hypothesis that Willman 1
density has remained constant throughout its evolution. Willman 1's
unusual kinematics challenges any current model of tidally disrupted
or ordered rotating system and point towards a significant tidal
evolution which could have stripped a noteworthy fraction of its
stellar component \citep{Willman:2010a}. The initial stellar density of the
system could have been such that binary systems were formed more
efficiently in the past. Consequently the probability estimate should
be taken {\it cum grano salis} since Willman 1 past evolution may have
played a crucial role in binary formation.  But since it might be
nearly impossible to derive the past encounter rate from current
dynamical structure, there is no reason to conclude that any
significant fraction of the point source population in the field is
associated with Willman 1.

\section{Summary and conclusions}
Adopting an estimated distance of 38 kpc, it might be the case that
the bulk of sources connected with Willman 1 lies below our luminosity
limit of $10^{32}$ erg s$^{-1}$ in the 0.5--2.0 keV energy band, as in
the case of globular cluster GLIMPSE-C01 \citep{Pooley:2007a}. However,
our encounter rate calculations indicate a very low probability of
finding candidate binaries within Willman 1. 
Combined with existing optical imaging, the \cha\ observation of Willman 1
 could help 
purge possible member stars
of any AGN contaminants/stellar interlopers. 
Furthermore, further analysis might help us investigate
how will the X-ray point source population 
influence the prospects of detecting a dark matter signal from 
Willman 1 with gamma-ray measurements.

Given the existing sterile neutrino 
limits, it might be wise to concentrate on the
detectability of the sterile neutrino with future X-ray experiments
\citep{Herder:2009a} as it will be difficult to improve current
measurements with the existing instrumentation \citep{Abazajian:2009b}. In
particular, making the case for future calorimeter experiments that
could definitely detect it or exclude it as a DM candidate
(F. Paerels, private communication).  We have shown here that any
viable indirect search for DM in this energy range must deal
not only with possible overlap with instrumental lines in the spectrum
\citep{RiemerSorensen:2006a}, but also with possible contamination from intervening
plasma lines that may mask a DM origin.  We caution that
extraordinary claims related to DM should be backed by
outstanding observational evidence \citep{Sagan:1980a}.

With the Large Hadron Collider (LHC) and other laboratory experiments
coming online, it might be possible to achieve a direct detection of
the particles responsible for DM.  However, any direct
detection in the laboratory will then have to be followed by
confirmation or dismissal of sameness in an astrophysical
context. Based on the present study, Willman 1 continues to hold a
select place among targets to conduct such searches. It is worth
noting that within the inner $5\arcmin$ radius circle of Willman 1,
there are no radio sources at 1.4 GHz in the NRAO VLA Sky Survey
(NVSS) source catalog \citep{Condon:1998a}.  Moreover, we have shown that
the point source population to a limiting 0.5--2.0 keV X-ray flux of
$6\times 10^{-16}$ erg cm$^{-2}$ s$^{-1}$ is consistent with
background AGN and/or foreground stars.  Pending 
a final verdict  regarding the kinematic distribution of Willman 1
\citep{Willman:2010a}, the available data from radio
through X-rays thus make Willman 1 a notable candidate for the
eventual astrophysical verification of a DM particle.

\section*{Acknowledgments}
We are indebted to an anonymous colleague for careful reading of the
original version.  We thank all the members of Grupo de Altas
Energ\'ias (GAE) at the Universidad Complutense de Madrid for
stimulating discussions.  N.M. acknowledges support from the Spanish
Ministry of Science and Innovation through a Ram\'on y Cajal
fellowship. We also acknowledge support from the Consolider-Ingenio
2010 Programme under grant MULTIDARK CSD2009-00064.

\bibliographystyle{JHEP}
\bibliography{mirabalnieto}

\end{document}